# Fluctuations of Electric Fields in the Active Site of the Enzyme Ketosteroid Isomerase

Valerie Vaissier Welborn[1,2,5,] and Teresa Head-Gordon[1,2,3,4,5,†]

[1]Kenneth S. Pitzer Center for Theoretical Chemistry, [2]Department of Chemistry, [3]Department of Bioengineering, [4]Department of Chemical and Biomolecular Engineering
University of California Berkeley
[5]Chemical Sciences Division, Lawrence Berkeley National Laboratory
Berkeley, California 94720, USA

[†]Corresponding author: thg@berkeley.edu

We report the effect of conformational dynamics on the fluctuations of electric fields in the active site of the enzyme Ketosteroid Isomerase (KSI). While KSI is considered rigid with little conformational variation to support different stages of the catalytic cycle, we show that KSI utilizes cooperative side chain motions of the entire protein scaffold outside the active site, which contribute negligibly to the electric fields on the substrate, by progressively stabilizing electric field contributions by particular active site residues at different timescales. The design of synthetic enzymes could benefit from strategies that can take advantage of the dynamics by using electric fields fluctuations as a guide.

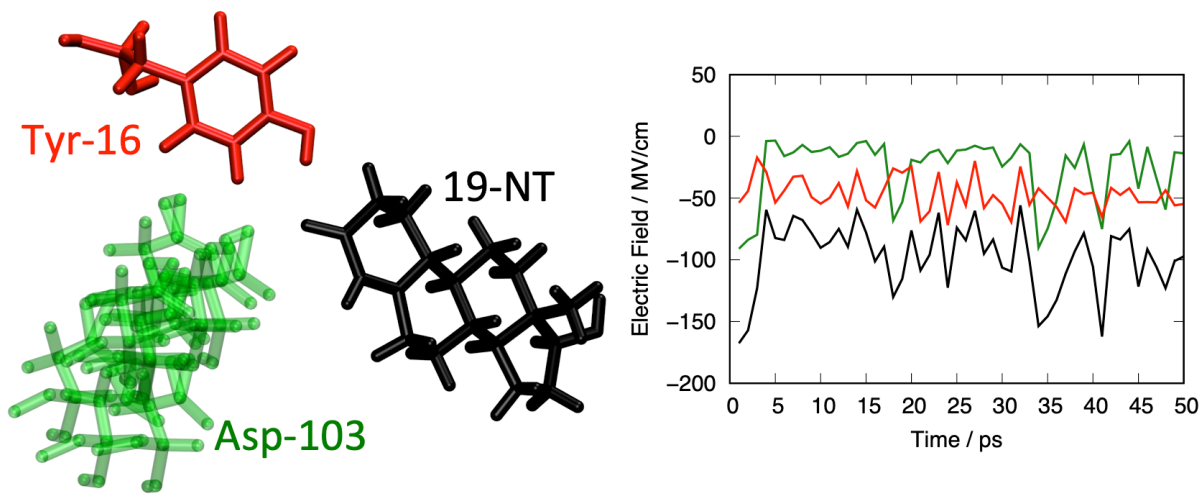

The field of enzyme catalysis is undergoing a very active phase to improve our understanding of enzymatic function in order to enhance synthetic enzyme design.[1] As for all enzymes, the detailed active site energetics plays a central role in the catalytic outcome, and its characterization has reached a stage where theoretical connections to experiment are tightly coupled. The pivotal role of electrostatics in the preorganized enzyme environment has long been established[2-4] because it eloquently serves as a unified descriptor of the molecular interactions that are responsible for the stabilization of the reaction transition state.[5] In fact our group successfully used electric fields to predict and ultimately improve enzymatic efficiency, effectively replacing the early rounds of laboratory directed evolution to increase enzyme performance by one to two orders of magnitude for a designed Kemp Eliminase[1, 6].

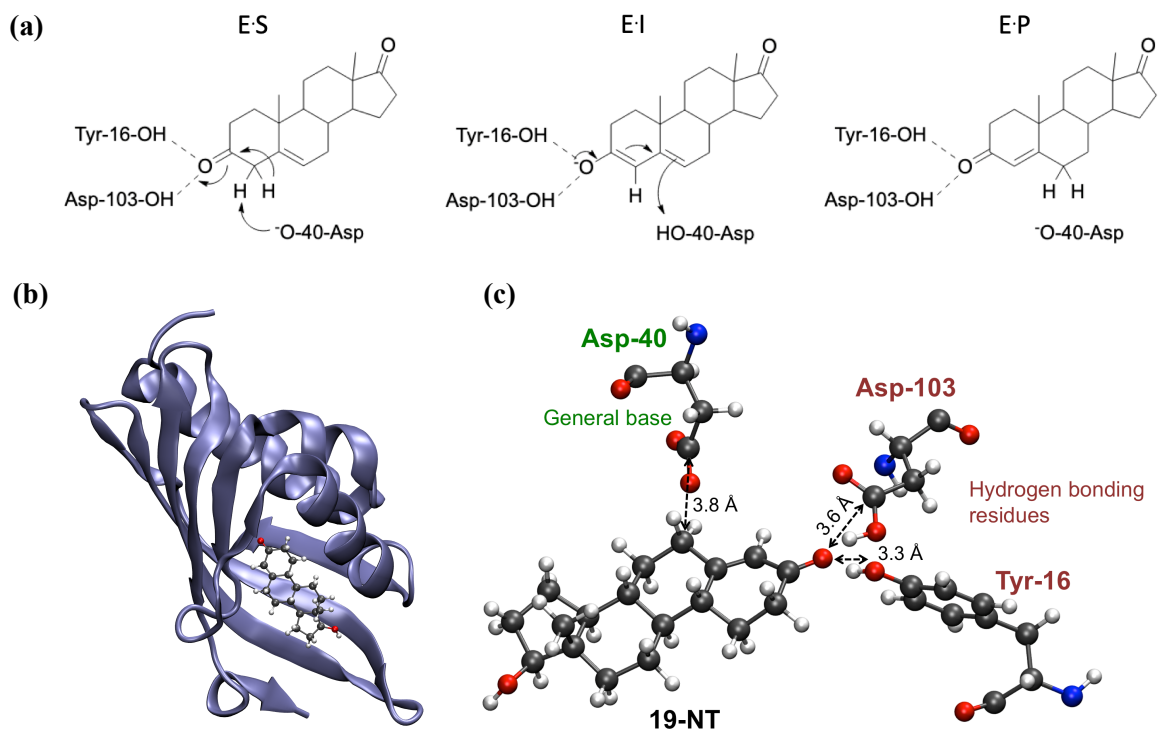

**Figure 1:** *Ketosteroid Isomerase.* (a) Reaction mechanism following a two-step acid/base process. (b) Enzyme in cartoon representation with the inhibitor 19-NT in the active site (c) 19-NT with the base Asp-40 and the two hydrogen bond making residues Asp-103 and Tyr-16. Color key: gray = carbon, red = oxygen, white = hydrogen, blue = nitrogen. PDB code: 5KP4.[7]

For KSI much work has been devoted to the energetics of the active site and in particular to the role of the extended hydrogen bond network that stabilizes the oxyanion hole.[8] A previous study found compelling evidence for long-range structural coupling within the hydrogen bond network of a group of tyrosines, with minimal influence of the remainder of the protein scaffold.[9]

In a QM/MM study, Hammes-Schiffer and colleagues identified water molecules with a direct role in the KSI active site, while also demonstrating that their presence did not conflict with the formation of short hydrogen bonds stabilizing the KSI intermediates.[10-11] Pioneered by Fried et al.[12-16], electric fields are becoming an increasingly popular tool to rationalize catalytic effects, especially in enzymes[1, 5-6]. They performed vibrational Stark Spectroscopy experiments using carbonyl probes to measure the electric fields in the active site KSI and KSI mutants,[5, 17] including conservative variants where tyrosines were replaced by 3-chlorocytosines.[7] From these experiments, they established a linear correlation between the strength of the electric fields in the active site and the activation free energy of the reaction, and a direct connection between the strength of a hydrogen bond and the electric field it exerts.[7] Subsequent theoretical studies have found support for the idea that electric fields strongly correlate with the efficiency of the catalytic process for KSI[18]. Markland and co-workers used *ab initio* path integral molecular dynamics (MD) simulations to further characterize KSI's active site hydrogen bond network,[19] observing enhanced proton flexibility within the active site. They also proposed an atomistic interpretation of the high electric fields in the active site of KSI[8] in which only a few first shell residues, namely Tyr-16 and Asp-103, were responsible for the fields, a result consistent with earlier work highlighting a very compact active site[20].

KSI has been traditionally categorized as a very stiff protein with limited catalytically relevant conformational changes.[5, 12] This viewpoint would be supported in Figure 2, which uses the AMOEBA polarizable force field to characterize the electric fields projected onto the carbonyl bond of 19-NT from water as well as a breakdown from different residues of KSI. Overall, the classical electrostatic field is calculated to be –108.9 ± –4.9 MV/cm, which is in reasonable agreement with the work of Fried et al., especially when considering the full range of uncertainty in the fitting procedure used to convert Stark spectroscopy measurements to electric field values (from –120 and –150 MV/cm).[17] Furthermore, we find that the majority of electric field effects comes from the three main catalytic residues: the general base, Asp-40, which contributes –15.85 MV/cm, and the residues Tyr-16 and Asp-103 which contribute –44.47 MV/cm and –37.75 MV/cm, respectively. This is in qualitative agreement with *ab initio* MD simulations that reported 95% of the electric field in the active site of KSI comes from Tyr-16 and Asp-103.[8] Previous theoretical studies have emphasized the need for polarization to compute electric fields[9, 21]; to further illustrate this point, our simulations ran without mutual polarization reduces the average electric field by 40 MV/cm (–108.93 ± –4.9 to –68.08 ± –3.1). Hence our calculations using full mutual polarization can capture nearly all the physics governing the production of electric fields in KSI, for which it has the undeniable advantage of accessing longer timescales relevant to this study.

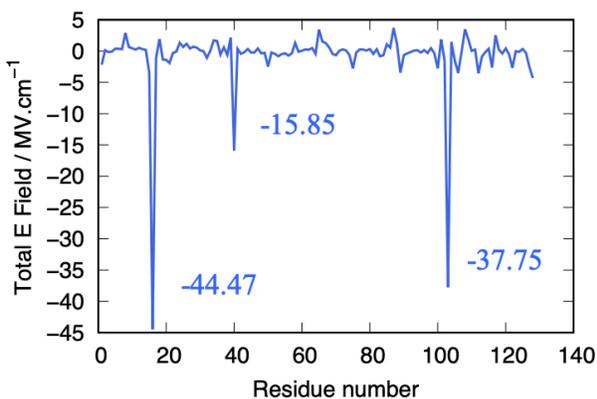

**Figure 2:** *Individual residue contribution to the electric field of KSI complexed with 19-NT.* Electric field projections on the carbonyl bond of the reactant analog 19-NT from each of the 128 residues. The last data point is the contribution of the water, which is the sum of active site (–9.46 MV/cm) and bulk (5.16 MV/cm) water contributions.

The results presented thus far are based on average values of the field, similarly to what can be found in previous literature on KSI.[5, 8, 18, 21] But because enzyme function is also governed by different time scales[22-24], then it might be expected that electric field fluctuations both in the active site and from the surrounding protein environment will modulate the enzyme activity.[25] Even rigid proteins undergo small backbone displacements[26-27] that in turn promote concerted side chain rearrangements[28-31] that can be important for catalytic function[32-33]. Zoi et al. showed that electric fields in the active site of KSI at the sub-picosecond timescale increases just before the first step of the isomerization reaction and decreases as it proceeds. But here we consider electric field changes as KSI undergoes conformational changes on much longer timescales than sub-picosecond vibrational motions, by simulating a combination of backbone variability[26] and side chain repacking[34] to access hundreds of nanosecond to microsecond timescales.[35-37] Our Monte Carlo Side Chain Entropy (MC-SCE) approach[34] has shown excellent agreement with experimental J-coupling measurements[38] across a wide selection of proteins and protein complexes that access side chain motions on these longer timescales, as well as confirming alternate side chain rotamers important for catalysis and allosteric regulation derived from both room temperature and cryogenically cooled X-ray crystallographic structures[32-33]. These conformational ensembles that represent the longer timescales are then coupled to the electric field calculation using the AMOEBA polarizable force field at the picosecond timescale.

Each panel of Figure 3 shows the underlying electric field fluctuations in the active site of KSI on the picosecond timescale, in which the electric fields fluctuate on the order of ± 20 MV/cm. Isolating the contribution from the key residues we find that these picosecond fluctuations come from the substrate hydrogen bonding residues Asp-103 (green) and Tyr-16 (red), driven by small fluctuations in bond length (below 0.5 Å) between them and the carbonyl group of the 19-NT substrate. By contrast, we observe almost no fluctuations in the electric field emanating from the Asp-40 catalytic base (orange), a signature that its position is optimized within the active site and largely impervious to motions on timescales greater than tens of picoseconds.

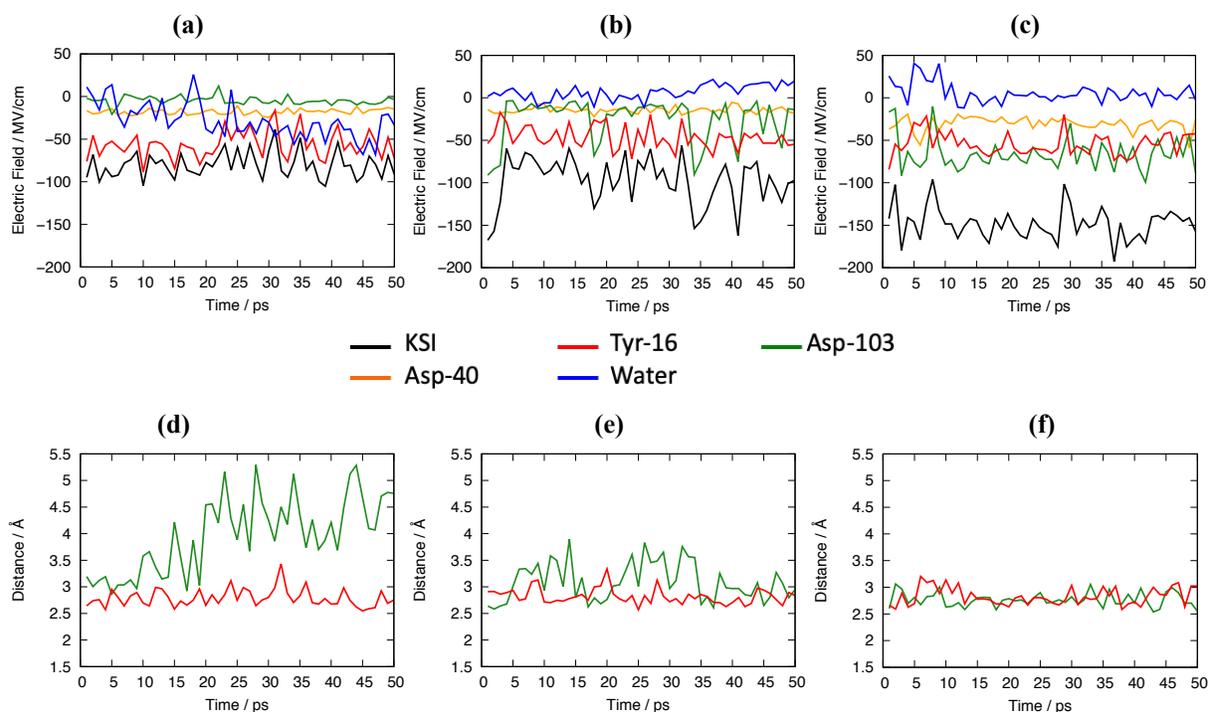

**Figure 3:** *Fluctuation of the electric field of KSI complexed with 19-NT over picoseconds to microseconds.* Panels (a-c) show the projected electric fields onto the 19-NT carbonyl bond emanating from the entire protein (black), residue Asp-40 (orange), residue Tyr-16 (red), residue Asp-103 (green) and water (blue). Panels (d-f) measure the corresponding distance between oxygens of Asp-103 (green) and Tyr-16 (red) to the 19-NT carbonyl oxygen. Each individual panel represents picosecond motions, whereas across the (a)-(c) and (d)-(f) panels are effectively on nanoseconds to millisecond timescales.

However, the averaging process on the picosecond timescale also obscures the underlying heterogeneity of the electric field data on even longer timescales, in which we find even larger changes of the average electric field depending on the underlying conformational state. This is demonstrated as we now compare panels across Figure 3(a)-(c), in which we observe over a 60 MV/cm change of the total electric field dependent on the wholesale side chain repacking on the longer timescale that seeded the picosecond trajectory. On this longer timescale the hydrogen-bonded network of the supporting tyrosines help anchor the position and orientation of Tyr-16 with respect to the substrate, and therefore its electrostatic field contribution is large but remains relatively constant over the longer nanosecond to microsecond timescale. But what varies most in the active site of KSI, and hence where most of the electric field variations come from over the longer timescale, is from Asp-103, which sensitively changes its hydrogen-bonding positioning to magnify or diminish its contribution on the total electric field, and hence the efficiency of the reaction (Figure 3(d)-(f)). These electric field changes correspond to the natural variations that would be expected from collective protein motions that reorients the Asp-103 residue on longer timescales such as the microsecond timescale of substrate turnover.[22, 39]

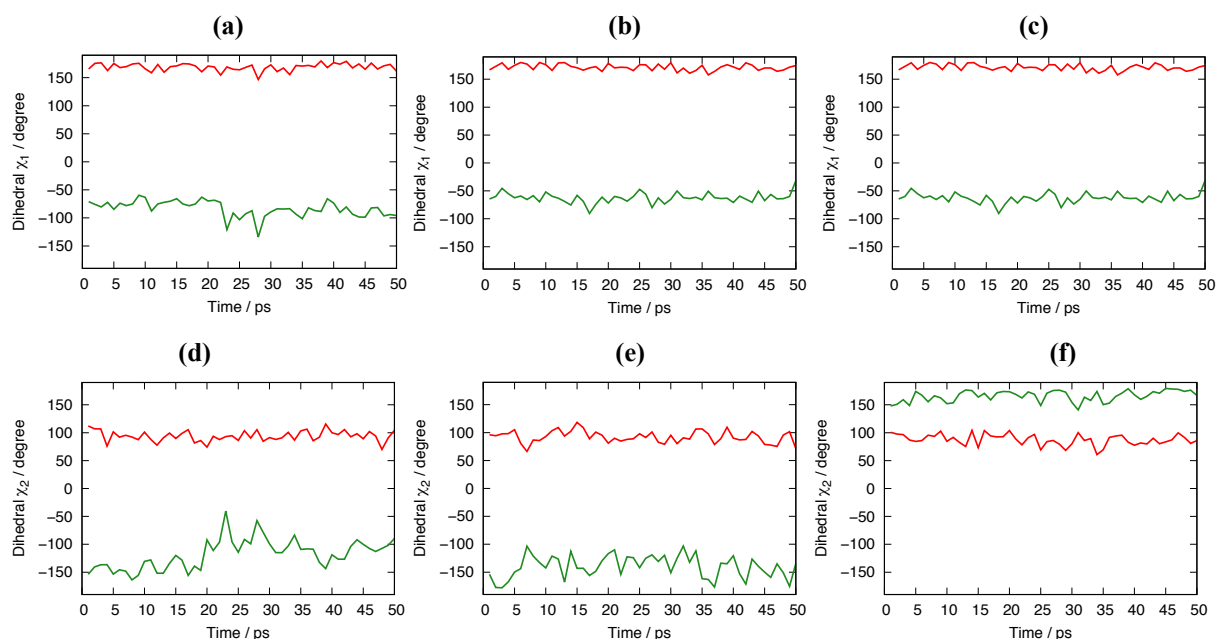

**Figure 4:** *Characterization of the rotamer states of Tyr-16 and Asp-103.* Panels (a-c) show the $\chi_1$ rotamer of Asp-103 (in green) and Tyr-16 (in red) associated to the electric fields in Figure 3a-c, while panels (d-f) show the $\chi_2$ rotamer.

Schwartz and co-workers have stated that, for KSI, "the same protein conformation is used for binding the substrate, isomerizing it, and allowing subsequent product release", whereas other enzymes rely on slower dynamics and hence alternative conformations to complete a catalytic cycle[25]. When considering the $\chi_1$ and $\chi_2$ rotamers states for Tyr-16 in Figure 4, it would support that view, since this side chain exhibits dihedral fluctuations of ± 10-20° that is consistent with a single well conformational state[40-41], and consistent with its static albeit large electric field contribution. However, the $\chi_1$ and $\chi_2$ rotamers states for Asp-103 show far more variation over the longer timescale (Figure 4), as large as ± 50-80°, indicating transitions between multiple dihedral angle wells that tracks much greater variability in its electric field contributions in the active site.

Thus KSI in fact acts as do other enzymes, but with functional motions that are more subtle – i.e. changing the rotamer state of a single active site residue – by raising the transition state barrier via an unfavorable electric field fluctuation on longer timescales to promote other functional states in the enzyme cycle, for example, substrate binding or product release. These findings are also consistent with a gating mechanism whereby the reaction can proceed at specific conformations where the key residue(s), in this case the Asp-103 residue, are optimally positioned[42]. This is likely one problem in enzyme design where often the active site is "overdesigned", for example the assumption that the role of Asp-103 is to stabilize the oxyanion hole, and thus fails to incorporate productive coupled motions on different timescales that supports the enzyme turnover cycle[43].

In summary, these results reconcile the idea that electric field fluctuations manifested through multi-timescale conformational dynamics are paramount to enzyme catalysis, just like energetic electrostatic stabilization of the transition state, by modulating the catalytically relevant conformations of the complete cycle[43]. This has several consequences starting with drawing attention to the importance of long-time sampling of conformational ensembles when interpreting experimental observables, as average fields from vibrational Stark Spectroscopy cannot be explained with single trajectory of tens of picoseconds. Second this emphasizes the role of dynamics in modulating the electric fields in the active site by magnifying or diminishing their contributions, and hence affecting the efficiency of the catalytic cycle. This is especially important to consider when designing new enzymes for which we can promote stable, catalytically relevant conformations on multiple timescales using electric fields fluctuations as a guide.

**METHODS**

*Sampling approaches to creating structural ensembles for KSI.* The KSI starting structure was found in the protein databank with docked 19-NT (PDB code : 5KP4[7]). We created 25 uncorrelated and low energy backbone structures using backrub simulations provided within the Rosetta modeling software package[26, 44-45]. To generate side chain ensemble diversity, we used our recently developed Monte Carlo method (MC-SCE)[34], for which we have performed extensive validation across ~60 proteins through comparison to high quality X-ray crystallography data and NMR experiments, to characterize the side chain structural ensemble on the 25 backbone structures for KSI. Briefly, the method uses a Rosenbluth side chain growth protocol sampled from an expanded side chain rotamer library[46-47], which are energy weighted according to a physical energy function based on the AMBERff99SB protein force field combined with a GB-HPMF implicit solvent model.[34]

*Electric field calculations.* Using Gromacs[48], the 25 lowest energy structures were explicitly solvated using a pre-equilibrated water box. We then performed 50 ps equilibration followed by 50 ps production runs in the NPT ensemble using the Tinker software package[49-51] and the AMOEBA polarizable force field[49, 52] to provide a high quality description of electrostatics in the active site and overall scaffold and solvent. We restrained 19-NT in place using 3 harmonic potentials with 1000 N/m spring constants, between the inhibitor and residues 16 (2.5 – 3.5 Å), 40 (3.5 – 4.5 Å) and 103 (5.5 – 6.5 Å). Electric fields are calculated every picosecond over the 50ps production run at the 2 atoms that make up the carbonyl bond. The electric field is projected onto each bond and the value reported is given by the mean of the electric field at the two atoms involved[53].

**ACKNOWLEDGMENTS.** This work was supported by the Director, Office of Science, Office of Basic Energy Sciences, Chemical Sciences Division of the U.S. Department of Energy under Contract No. DE-AC02-05CH11231. This research used resources of the National Energy Research Scientific Computing Center, a DOE Office of Science User Facility supported by the Office of Science of the U.S. Department of Energy under Contract No. DE-AC02-05CH11231.

**REFERENCES**

1. Vaissier Welborn, V.; Head-Gordon, T., Computational Design of Synthetic Enzymes. *Chem. Rev.* **2018**, *in press*.
2. Warshel, A.; Sharma, P. K.; Kato, M.; Xiang, Y.; Liu, H.; Olsson, M. H., Electrostatic basis for enzyme catalysis. *Chem. Rev.* **2006**, *106* (8), 3210-35.
3. Warshel, A., Energetics of enzyme catalysis. *Proc. Natl. Acad. Sci. USA* **1978**, *75*, 5250-5254.
4. Warshel, A.; Weiss, R. M., An Empirical Valence Bond Approach for Comparing Reactions in Solutions and in Enzymes *J. Am. Chem. Soc.* **1980**, *102*, 6218-6226.
5. Fried, S. D.; Boxer, S. G., Electric Fields and Enzyme Catalysis. *Ann. Rev. Biochem.* **2017**, *86*, 387-415.

6.	Welborn, V. V.; Ruiz Pestana, L.; Head-Gordon, T., Computational optimization of electric fields for better catalysis design. *Nature Catalysis* **2018,** *1* (9), 649-655.
7.	Wu, Y.; Boxer, S. G., A Critical Test of the Electrostatic Contribution to Catalysis with Noncanonical Amino Acids in Ketosteroid Isomerase. *J. Am. Chem. Soc.* **2016,** *138* (36), 11890-5.
8.	Wang, L.; Fried, S. D.; Markland, T. E., Proton Network Flexibility Enables Robustness and Large Electric Fields in the Ketosteroid Isomerase Active Site. *J. Phys. Chem. B* **2017,** *121* (42), 9807-9815.
9.	Pinney, M. M.; Natarajan, A.; Yabukarski, F.; Sanchez, D. M.; Liu, F.; Liang, R.; Doukov, T.; Schwans, J. P.; Martinez, T. J.; Herschlag, D., Structural Coupling Throughout the Active Site Hydrogen Bond Networks of Ketosteroid Isomerase and Photoactive Yellow Protein. *J. Am. Chem. Soc.* **2018,** *140* (31), 9827-9843.
10.	Hanoian, P.; Hammes-Schiffer, S., Water in the active site of ketosteroid isomerase. *Biochem.* **2011,** *50* (31), 6689-700.
11.	Layfield, J. P.; Hammes-Schiffer, S., Calculation of vibrational shifts of nitrile probes in the active site of ketosteroid isomerase upon ligand binding. *J. Am. Chem. Soc.* **2013,** *135* (2), 717-25.
12.	Sigala, P. A.; Fafarman, A. T.; Schwans, J. P.; Fried, S. D.; Fenn, T. D.; Caaveiro, J. M.; Pybus, B.; Ringe, D.; Petsko, G. A.; Boxer, S. G.; Herschlag, D., Quantitative dissection of hydrogen bond-mediated proton transfer in the ketosteroid isomerase active site. *Proc. Natl. Acad. Sci. USA* **2013,** *110* (28), E2552-61.
13.	Fafarman, A. T.; Sigala, P. A.; Schwans, J. P.; Fenn, T. D.; Herschlag, D.; Boxer, S. G., Quantitative, directional measurement of electric field heterogeneity in the active site of ketosteroid isomerase. *Proc. Natl. Acad. Sci. USA* **2012,** *109* (6), E299-308.
14.	Fafarman, A. T.; Sigala, P. A.; Herschlag, D.; Boxer, S. G., Decomposition of vibrational shifts of nitriles into electrostatic and hydrogen-bonding effects. *J. Am. Chem. Soc.* **2010,** *132* (37), 12811-3.
15.	Fafarman, A. T.; Webb, L. J.; Chuang, J. I.; Boxer, S. G., Site-specific conversion of cysteine thiols into thiocyanate creates an IR probe for electric fields in proteins. *J. Am. Chem. Soc.* **2006,** *128* (41), 13356-7.
16.	Jha, S. K.; Ji, M.; Gaffney, K. J.; Boxer, S. G., Direct measurement of the protein response to an electrostatic perturbation that mimics the catalytic cycle in ketosteroid isomerase. *Proc. Natl. Acad. Sci. USA* **2011,** *108* (40), 16612-7.
17.	Fried, S. D.; Bagchi, S.; Boxer, S. G., Extreme electric fields power catalysis in the active site of ketosteroid isomerase. *Science* **2014,** *346* (6216), 1510-4.
18.	Wang, X.; He, X., An Ab Initio QM/MM Study of the Electrostatic Contribution to Catalysis in the Active Site of Ketosteroid Isomerase. *Molecules* **2018,** *23* (10).
19.	Wang, L.; Fried, S. D.; Boxer, S. G.; Markland, T. E., Quantum delocalization of protons in the hydrogen-bond network of an enzyme active site. *Proc. Natl. Acad. Sci. USA* **2014,** *111* (52), 18454-9.
20.	Somarowthu, S.; Brodkin, H. R.; D'Aquino, J. A.; Ringe, D.; Ondrechen, M. J.; Beuning, P. J., A tale of two isomerases: compact versus extended active sites in ketosteroid isomerase and phosphoglucose isomerase. *Biochem.* **2011,** *50* (43), 9283-95.
21.	Fried, S. D.; Wang, L. P.; Boxer, S. G.; Ren, P.; Pande, V. S., Calculations of the electric fields in liquid solutions. *J. Phys. Chem. B* **2013,** *117* (50), 16236-48.

22. Eisenmesser, E. Z.; Millet, O.; Labeikovsky, W.; Korzhnev, D. M.; Wolf-Watz, M.; Bosco, D. A.; Skalicky, J. J.; Kay, L. E.; Kern, D., Intrinsic dynamics of an enzyme underlies catalysis. *Nature* **2005,** *438* (7064), 117-21.
23. Henzler-Wildman, K. A.; Lei, M.; Thai, V.; Kerns, S. J.; Karplus, M.; Kern, D., A hierarchy of timescales in protein dynamics is linked to enzyme catalysis. *Nature* **2007,** *450* (7171), 913-6.
24. Henzler-Wildman, K.; Kern, D., Dynamic personalities of proteins. *Nature* **2007,** *450* (7172), 964-72.
25. Zoi, I.; Antoniou, D.; Schwartz, S. D., Electric Fields and Fast Protein Dynamics in Enzymes. *J. Phys. Chem. Lett.* **2017,** *8* (24), 6165-6170.
26. Friedland, G. D.; Linares, A. J.; Smith, C. a.; Kortemme, T., A simple model of backbone flexibility improves modeling of side-chain conformational variability. *J. Mol. Bio.* **2008,** *380*, 757-74.
27. Kohn, J. E.; Afonine, P. V.; Ruscio, J. Z.; Adams, P. D.; Head-Gordon, T., Evidence of functional protein dynamics from X-ray crystallographic ensembles. *PLoS Comp. Bio.* **2010,** *6* (8).
28. Moorman, V. R.; Valentine, K. G.; Wand, a. J., The dynamical response of hen egg white lysozyme to the binding of a carbohydrate ligand. *Prot. Sci.* **2012,** *21*, 1066-73.
29. Schnell, J. R.; Dyson, H. J.; Wright, P. E., Effect of cofactor binding and loop conformation on side chain methyl dynamics in dihydrofolate reductase. *Biochem.* **2004,** *43*, 374-83.
30. Tzeng, S.-R.; Kalodimos, C. G., Protein activity regulation by conformational entropy. *Nature* **2012,** *488*, 236-40.
31. Fenwick, R. B.; van den Bedem, H.; Fraser, J. S.; Wright, P. E., Integrated description of protein dynamics from room-temperature X-ray crystallography and NMR. *Proc. Natl. Acad. Sci. USA* **2014,** *111* (4), E445-54.
32. Fraser, J. S.; Clarkson, M. W.; Degnan, S. C.; Erion, R.; Kern, D.; Alber, T., Hidden alternative structures of proline isomerase essential for catalysis. *Nature* **2009,** *462*, 669-73.
33. Fraser, J. S.; van den Bedem, H.; Samelson, A. J.; Lang, P. T.; Holton, J. M.; Echols, N.; Alber, T., Accessing protein conformational ensembles using room-temperature X-ray crystallography. *Proc. Natl. Acad. Sci. USA* **2011,** *108* (39), 16247-52.
34. Bhowmick, A.; Head-Gordon, T., A Monte Carlo Method for Generating Side Chain Structural Ensembles. *Structure* **2015,** *23* (1), 44-55.
35. Smith, C. A.; Ban, D.; Pratihar, S.; Giller, K.; Schwiegk, C.; de Groot, B. L.; Becker, S.; Griesinger, C.; Lee, D., Population Shuffling of Protein Conformations. *Angew. Chem. Intl. Ed.* **2015,** *54* (1), 207-210.
36. Keedy, D. A.; Kenner, L. R.; Warkentin, M.; Woldeyes, R. A.; Hopkins, J. B.; Thompson, M. C.; Brewster, A. S.; Van Benschoten, A. H.; Baxter, E. L.; Uervirojnangkoorn, M.; McPhillips, S. E.; Song, J.; Alonso-Mori, R.; Holton, J. M.; Weis, W. I.; Brunger, A. T.; Soltis, S. M.; Lemke, H.; Gonzalez, A.; Sauter, N. K.; Cohen, A. E.; van den Bedem, H.; Thorne, R. E.; Fraser, J. S., Mapping the conformational landscape of a dynamic enzyme by multitemperature and XFEL crystallography. *eLife* **2015,** *4*, e07574.
37. Chou, J. J.; Case, D. A.; Bax, A., Insights into the Mobility of Methyl-Bearing Side Chains in Proteins from 3JCC and 3JCN Couplings. *J. Am. Chem. Soc.* **2003,** *125* (29), 8959-8966.

38. Tuttle, L. M.; Dyson, H. J.; Wright, P. E., Side-Chain Conformational Heterogeneity of Intermediates in the Escherichia coli Dihydrofolate Reductase Catalytic Cycle. *Biochem.* **2013,** *52* (20), 3464-3477.
39. Eisenmesser, E. Z.; Bosco, D. A.; Akke, M.; Kern, D., Enzyme dynamics during catalysis. *Science* **2002,** *295* (5559), 1520-3.
40. Li, F.; Grishaev, A.; Ying, J.; Bax, A., Side Chain Conformational Distributions of a Small Protein Derived from Model-Free Analysis of a Large Set of Residual Dipolar Couplings. *J. Am. Chem. Soc.* **2015,** *137* (46), 14798-14811.
41. Hansen, D. F.; Kay, L. E., Determining Valine Side-Chain Rotamer Conformations in Proteins from Methyl 13C Chemical Shifts: Application to the 360 kDa Half-Proteasome. *J. Am. Chem. Soc.* **2011,** *133* (21), 8272-8281.
42. Santos-Martins, D.; Calixto, A. R.; Fernandes, P. A.; Ramos, M. J., A Buried Water Molecule Influences Reactivity in α-Amylase on a Subnanosecond Time Scale. *ACS Catal.* **2018,** *8* (5), 4055-4063.
43. Nashine, V. C.; Hammes-Schiffer, S.; Benkovic, S. J., Coupled motions in enzyme catalysis. *Curr. Opin. Chem. Bio.* **2010,** *14* (5), 644-651.
44. Friedland, G. D.; Kortemme, T., Designing ensembles in conformational and sequence space to characterize and engineer proteins. *Curr. Opin. Struct. Bio.* **2010,** *20*, 377-84.
45. Friedland, G. D.; Lakomek, N.-A.; Griesinger, C.; Meiler, J.; Kortemme, T., A correspondence between solution-state dynamics of an individual protein and the sequence and conformational diversity of its family. *PLoS Comp. Bio.* **2009,** *5*, e1000393.
46. Shapovalov, M. V.; Dunbrack, R. L., A smoothed backbone-dependent rotamer library for proteins derived from adaptive kernel density estimates and regressions. *Structure* **2011,** *19*, 844-58.
47. Dunbrack, R. L., Rotamer Libraries in the 21st Century. *Curr. Opin. Struct. Bio.* **2002,** *12* (4), 431-440.
48. Abraham, M. J.; Murtola, T.; Schulz, R.; Páll, S.; Smith, J. C.; Hess, B.; Lindahl, E., GROMACS: High performance molecular simulations through multi-level parallelism from laptops to supercomputers. *SoftwareX* **2015,** *1–2*, 19-25.
49. Ponder, J. W.; Wu, C.; Ren, P.; Pande, V. S.; Chodera, J. D.; Schnieders, M. J.; Haque, I.; Mobley, D. L.; Lambrecht, D. S.; DiStasio, R. A., Jr.; Head-Gordon, M.; Clark, G. N.; Johnson, M. E.; Head-Gordon, T., Current status of the AMOEBA polarizable force field. *J. Phys. Chem. B* **2010,** *114* (8), 2549-64.
50. Ponder, J. W. *Tinker--Software Tools for Molecular Design*, 7.0; 2014.
51. Albaugh, A.; Boateng, H. A.; Bradshaw, R. T.; Demerdash, O. N.; Dziedzic, J.; Mao, Y.; Margul, D. T.; Swails, J.; Zeng, Q.; Case, D. A.; Eastman, P.; Wang, L.-P.; Essex, J. W.; Head-Gordon, M.; Pande, V. S.; Ponder, J. W.; Shao, Y.; Skylaris, C.-K.; Todorov, I. T.; Tuckerman, M. E.; Head-Gordon, T., Advanced Potential Energy Surfaces for Molecular Simulation. *J. Phys. Chem. B* **2016,** *120* (37), 9811-9832.
52. Ren, P.; Wu, C.; Ponder, J. W., Polarizable Atomic Multipole-based Molecular Mechanics for Organic Molecules. *J. Chem. Theo. Comp.* **2011,** *7* (10), 3143-3161.
53. Bhowmick, A.; Sharma, S. C.; Head-Gordon, T., The Importance of the Scaffold for de Novo Enzymes: A Case Study with Kemp Eliminase. *J. Am. Chem. Soc.* **2017,** *139* (16), 5793-5800.